\documentclass[aps,prb,reprint,showpacs]{revtex4-2}
\usepackage{graphicx} 
\graphicspath{{Images/}}
\usepackage{amsmath} 
\usepackage{mathtools} 
\usepackage{gensymb} 
\usepackage{textcomp}
\usepackage{bm} 
\usepackage{natbib}
\usepackage{natmove}
\usepackage[colorlinks=true, linkcolor=blue, citecolor=blue, urlcolor=blue, linktoc=page, bookmarks=false, pdfstartview={FitH}, pdfborder={0 0 0.0 [3 3]}]{hyperref} 
\usepackage{color} 
\usepackage{dcolumn} 
\newcolumntype{.}{D{.}{.}{2.1}}
\newcolumntype{-}{D{.}{.}{4.0}}
\usepackage{makecell}
\usepackage{multirow}
\usepackage{cleveref}
\usepackage{easyReview} 
\crefname{figure}{Fig.}{Figs}
\Crefname{figure}{Figure}{Figures}
\crefname{table}{Table}{Tables}
\crefname{equation}{Eq.}{Eqs.}
\crefname{section}{Sec.}{Secs.}
\renewcommand{\today}{\number\day \space \ifcase \month \or January\or February\or March\or April\or May\or June\or July\or August\or September\or October\or November\or December\fi \space \number\year} 
\begin{document}
\title{Proximity-induced Rashba spin-orbit interaction in BaMnO$_\text{3}|$KTaO$_\text{3}$ heterostructure for antiferromagnetic spintronics}
\author{Vivek \surname{Kumar}}
\author{Nirmal \surname{Ganguli}}
\email[Contact author: ]{NGanguli@iiserb.ac.in}
\affiliation{Department of Physics, \href{https://ror.org/02rb21j89}{Indian Institute of Science Education and Research Bhopal}, Bhauri, Bhopal 462066, India}
\date{\today}
\begin{abstract}
Antiferromagnetic spintronics, a promising technology for ultra-fast electronic devices, faces several challenges, including the lack of materials simultaneously hosting robust antiferromagnetism and adequate Rashba-like interaction. We design a heterostructure of BaMnO$_3|$KTaO$_3$ with the idea of proximity-inducing moderate Rashba spin-orbit interaction from KTaO$_3$ part to BaMnO$_3$ part, where the latter is already a robust antiferromagnet. Within our DFT calculations, the heterostructure reveals BaMnO$_3$ bands near the Fermi level with a significant magnetic moment per Mn atom and a decent ordering temperature. Further, the BaMnO$_3$ bands in the heterostructure exhibit linear Rashba interaction with a sizable Rashba coefficient, owing to its proximity to KTaO$_3$. Our work can motivate future research by demonstrating the road map to proximity-induced Rashba interaction for antiferromagnetic spintronics.
\end{abstract}
\maketitle

\section{\label{sec:intro}Introduction}
Spin-orbit interaction (SOI) partnered with inversion asymmetry is known to result in spin splitting and helical spin textures that may help realize tunable spin-orbit torque, a key to antiferromagnetic spintronics \cite{BaltzRMP18}. However, simultaneously realizing robust antiferromagnetism and moderately strong SOI has been a challenge, slowing the progress in this area. Oxide heterostructures host several attractive features ranging from a two-dimensional electron system to magnetism and Rashba interaction \cite{OhtomoN04, GanguliPRL14, BertNP11, LiNP11, ZhongPRB13, ChakrabortyPRB20, VicenteAM21, KumarPRB22}, making them excellent testbeds for future technology. \citet{ZhongPRB13} predicted a Rashba interaction in LaAlO$_3|$SrTiO$_3$ heterostructure, for which \citet{LinNC19} estimated a weak Rashba coefficient. KTaO$_3$ (KTO) (001) surface shows a giant Rashba splitting \cite{VarottoNC22, GuptaAM22, SetvinS18, KumarPRB22, KimPRB14, OjhaPRB21}, although it hosts no magnetism. Earlier, we designed and simulated a heterostructure of LaAlO$_3|$SrIrO$_3|$SrTiO$_3$ to be a suitable platform for antiferromagnetic spintronics \cite{ChakrabortyPRB20, GanguliPRB25}, where we observed moderate Rashba-Dresselhaus interaction, although a weak antiferromagnetism with a small magnetic moment coming from the Ir-$5d$ states.

We propose an oxide heterostructure to find an optimum solution to this problem using proximity effect, where some features of one component of a heterostructure, including magnetism, superconductivity, and spin-orbit interaction, are induced to the other component, making the heterostructure host features far more promising for technology than the collection of the individual components' features \cite{MarchenkoNC12, EremeevPRB13, AvsarNC14, QiaoPRL14, KatmisN16, IslandN19, TamuraPRB19, TongPRAp19, LopezPRB19}. In an earlier work on LaAlO$_3|$SrIrO$_3|$SrTiO$_3$ heterostructure, we observed stronger than usual spin splitting of the Ti-$3d$ states besides a moderate Rashba-Dresselhaus interaction in the Ir-$5d$ states \cite{ChakrabortyPRB20}. Motivated by this observation, here we consider a heterostructure of KTO, a perovskite oxide showing strong Rashba interaction owing to Ta-$5d$ states, with BaMnO$_3$ (BMO), a robust antiferromagnet owing to the Mn-$3d$ states \cite{RondinelliPRB09, SatapathyAPL12}. Although most stable in a hexagonal structure \cite{ChamberlandJSSC70}, a cubic phase of BMO can be realized \cite{GokogluCMS11}. Thus, when grown on a thick substrate of cubic KTO, BMO is expected to grow in a cubic phase with a reasonable lattice match \cite{SondenaPRB07}.

In this work, we simulate a computationally tractable BMO$|$KTO heterostructure where the BMO component exhibits antiferromagnetic order up to a moderate temperature. The electronic bands are found to align in a type-I fashion, where both valence band maximum and conduction band minimum correspond to the BMO part, owing to the electrostatic potential development, as discussed below.
Further, our calculations reveal that a moderate Rashba spin-orbit interaction can be induced in the BMO part, paving the way to overcome many pragmatic challenges for technology based on antiferromagnetic spintronics. The remainder of the article have been organized as follows: \cref{sec:Method} describes the methodology adopted for our calculations. The electrostatic potential developed due to heterostructuring of a polar and a non-polar material, leading to a possible charge transfer is elaborated in \cref{sec:potential}. Our results are discussed in \cref{sec:Result}. Finally, we summarize and conclude the article in \cref{sec:conc}.
\section{\label{sec:Method}Methodology}
We simulate a heterostructure with a periodically repeated supercell comprising three unit cell thick KTO and two unit cell thick BMO perpendicular to the interface, henceforth denoted as the (BaMnO$_3$)$_2|$(KTaO$_3$)$_3$ heterostructure. Assuming epitaxial growth of BMO on a thick substrate of KTO, we fix the heterostructure's in-plane lattice constant to 3.988~\AA\ matching that of KTO \cite{KTOlatticeConstant}, and allow optimization for the lattice constant along the $c$-direction. We employed density functional theory as implemented in the {\scshape vasp} code \cite{vasp1, vasp2} for the total energy, electronic structure, and magnetic properties calculations. Relativistic spin-orbit interaction (SOI) is considered for all calculations except structure optimization. Local spin density approximation (LSDA) \cite{ldaCA, PerdewPRB81} with Hubbard-$U$ correction \cite{DudarevPRB98} for strong electron-electron correction with $U_\text{eff} = U - J = 1$ and 3~eV for Ta-$5d$ and Mn-$3d$ states, respectively, is employed for describing the exchange-correlation functional. The projector augmented wave (PAW) method \cite{paw}, combined with a plane wave basis set with a cutoff energy of 450~eV, describes the potential and expands the wavefunctions. The Brillouin zone integrations are performed using the improved tetrahedron method \cite{BlochlPRB94T} over a $\Gamma$-centered $11 \times 11 \times 1$ $k$-point mesh. A small energy threshold of $10^{-7}$~eV is used for self-consistent calculations considering SOI. The atomic positions are optimized by minimizing the Hellman-Feynman forces on each atom to a tolerance of $10^{-2}$~eV\AA$^{-1}$.
\section{\label{sec:potential}Electrostatic potential}
The heterojunction between a polar and a nonpolar material experiences a charge transfer that plays a crucial role in its electronic properties. While the mechanism of charge transfer at perovskite oxide heterojunctions is debated, the electrostatic origin of the phenomenon is established \cite{GanguliPRL14, KumarPRB22}. Being a heterostructure of a nonpolar and a polar material, the electrostatic potential in BMO$|$KTO heterostructure will show similar patterns as other oxide heterostructures \cite{GanguliPRL14}.
\begin{figure}
    \includegraphics[scale = 0.33]{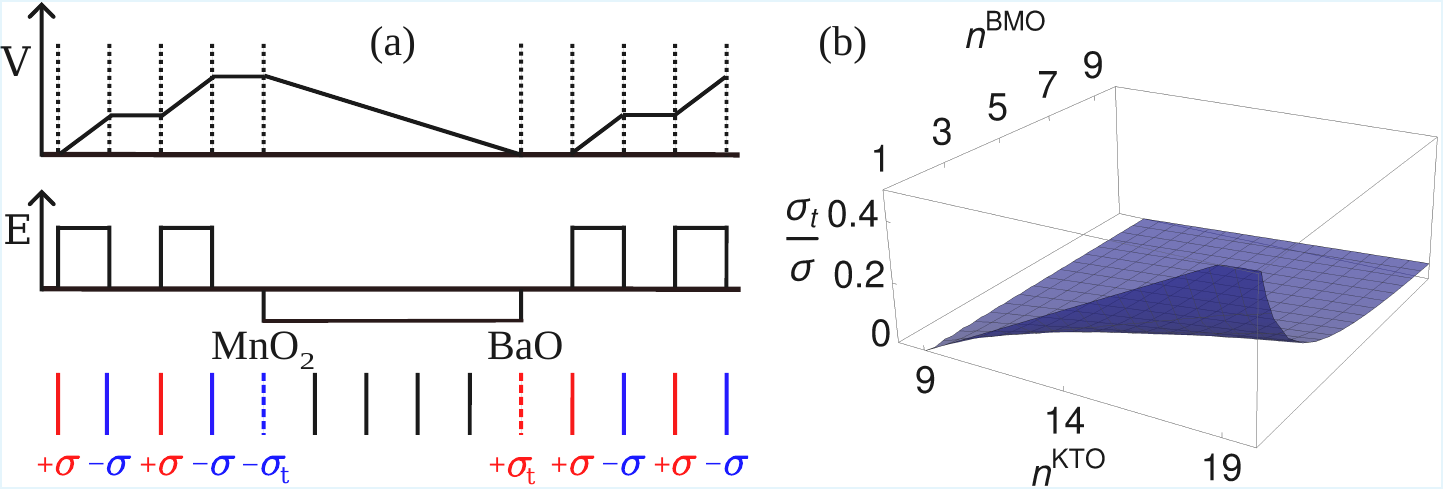}
    \caption{\label{fig:chargeTransfer}The electric field $E$ and the electrostatic potential energy $V$ due to the BMO$|$KTO superlattice are illustrated in (a) for the alternating $+1|-1$ charged planes in KTO and neutral planes in BMO, assuming a parallel plate capacitor model. The required charge transfer with increasing thickness of BMO and KTO layers is displayed in (b).}
\end{figure}
If a film of BMO is deposited on a thick KTO substrate, the electrostatic potential development may be described similarly to Refs.~\cite{GanguliPRL14, KumarPRB22}. However, the BMO$|$KTO superlattice considered here is distinct in its geometry; therefore, it calls for a compatible theoretical model describing the electrostatic potential. As illustrated in \cref{fig:chargeTransfer}(a) (bottom panel), we consider $n^\text{KTO}$ number of KTO unit cells, followed by $n^\text{BMO}$ number of BMO unit cells, and the arrangement repeats itself within periodic boundary conditions. Considering a heterostructure of the (001) planes, a KTO unit cell nominally comprises a (Ta$^{5+}$O$_2^{2-}$)$^+$ plane and a (K$^+$O$^{2-}$)$^-$ plane, resulting in an alternating arrangement of planes with $+\sigma$ and $-\sigma$  surface charge densities. On the other hand, a BMO unit cell nominally comprises a (Mn$^{4+}$O$_2^{2-}$)$^0$ plane and a (Ba$^{2+}$O$_2^{2-}$)$^0$ plane, both being charge neutral. The arrangement yields two heterojunctions: one between (KO)$^-$ and (MnO$_2$)$^0$ planes, and the other between (BaO)$^0$ and (TaO$_2$)$^+$ planes, as illustrated in \cref{fig:chargeTransfer}(a). The corresponding electrostatic potential energy would diverge for a large thickness of the constituent films, unless some charge transfer takes place to the MnO$_2$ and the BaO planes at the interfaces, leading to $-\sigma_t$ and $+\sigma_t$ surface charge densities, respectively, to contain the potential within a threshold value $V_\text{Th}$. As our calculations reveal a type-I band alignment where the valence band maximum and the conduction band minimum of the smaller band gap material BMO lies within the band gap of KTO, we expect the threshold potential $V_\text{Th}$ to be the sum of BMO's band gap $\varepsilon_g^\text{BMO}$ and the valence band offset $\Delta$. 
\begin{equation}
    V_\text{Th} = \varepsilon_g^\text{BMO} + \Delta \geq \frac{n^\text{KTO} a^\text{KTO} \sigma}{2 \epsilon^\text{KTO}} - \frac{n^\text{BMO} a^\text{BMO} \sigma_t}{\epsilon^\text{BMO}}, \label{eq:Potential}
\end{equation}
where $\epsilon^\text{KTO}$ and $\epsilon^\text{BMO}$ are the permittivity of KTO and BMO, respectively, while $a^\text{KTO}$ and $a^\text{BMO}$ denote the inter-planar distance in KTO and BMO parts of the heterostructure. The first term in the right-hand side of \cref{eq:Potential} corresponds to the uprising lines in the potential energy in the KTO part (see \cref{fig:chargeTransfer}(a)), while the second term represents the downward line in the BMO part. We may express the ratio of transferred surface charge density to the KTO surface charge density as
\begin{equation}
    \frac{\sigma_t}{\sigma} \geq \frac{\epsilon^\text{BMO}}{n^\text{BMO} a^\text{BMO}} \left( \frac{n^\text{KTO} a^\text{KTO}}{2 \epsilon^\text{KTO}} - \frac{V_\text{Th}}{\sigma} \right). \label{eq:ChargeTransfer}
\end{equation}
The model suggests that a charge transfer to the interface is required for
\begin{equation}
    n^\text{KTO} > \frac{2 \epsilon^\text{KTO} V_\text{Th}}{a^\text{KTO} \sigma}. \label{eq:TransferCondition}
\end{equation}
We gather from \cref{eq:ChargeTransfer,eq:TransferCondition} that the onset of charge transfer at the interface depends only on the KTO thickness in the superlattice. In contrast, BMO thickness impacts the amount of charge transfer once the onset condition is met. Using a set of reasonable parameter values \footnote{We use $\epsilon^\text{KTO} = 220 \epsilon_0$ \cite{AgrawalJPCSSP70}, $\epsilon_0$ being the free-space permittivity, $\epsilon^\text{BMO} = 22 \epsilon_0$ \cite{HughesAFM24}, $\varepsilon_g^\text{BMO} = 1.08$~eV \cite{GokogluCMS11}, $\Delta = -0.145$~eV, and $a^\text{KTO} = 3.988$~\AA\ for the calculations.}, we estimate that 10-unit-cell-thick KTO is required for the onset of a charge transfer. \Cref{fig:chargeTransfer}(b) displays the ratio of transferred surface charge density to the value of surface charge density in KO or TaO$_2$ planes as a function of the KTO and BMO thickness, illustrating the charge transfer landscape in the system. The figure suggests a required onset of charge transfer with at least 10 unit cell thick KTO and more charge transfer with increasing KTO thickness, while a thicker BMO layer would attenuate the charge transfer. However, unlike the KTaO$_3$ slab \cite{KumarPRB22} or LaAlO$_3|$SrTiO$_3$ heterostructure \cite{GanguliPRL14}, \cref{eq:ChargeTransfer} does not suggest any asymptotic limit for $\sigma_t/\sigma$ in this superlattice; the ratio monotonically increases with increasing KTO thickness. Simulating different antiferromagnetic arrangements requires a $2a \times 2b$ in-plane breadth. To make our calculations tractable, we limit the thickness of BMO and KTO layers to 2 and 3 unit cells, respectively. Although the above discussion does not warrant a charge transfer in this limit, we can investigate the crucial magnetic properties and Rashba interaction within this computationally tractable simulation cell.
\section{\label{sec:Result}Results and Discussion}
Since the hexagonal structure of BaMnO$_3$ is the most stable one, we begin by assessing the static and dynamic stability of the cubic BaMnO$_3$ structure considered here as a component of the heterostructure. We check the mechanical stability by calculating the elastic tensor and find the structure to be stable in its cubic form \cite{HuAMSE18}. Further, calculating the phonon dispersion within density functional perturbation theory as implemented in {\scshape vasp}, we find the structure to be dynamically stable, since no imaginary phonon frequency is observed. After ensuring the static and dynamic stability of cubic BaMnO$_3$, we start with simulating a $2a \times 2b$ cell for (BaMnO$_3$)$_2|$(KTaO$_3$)$_3$ heterostructure within periodic boundary conditions, leading to a supercell arrangement. Unlike in Ref.~\cite{GanguliPRB25}, here we choose an inversion-asymmetric structure to allow for the possibility of Rashba-like interaction in the absence of any structural distortion.
\begin{figure*}
\includegraphics[scale = 0.185]{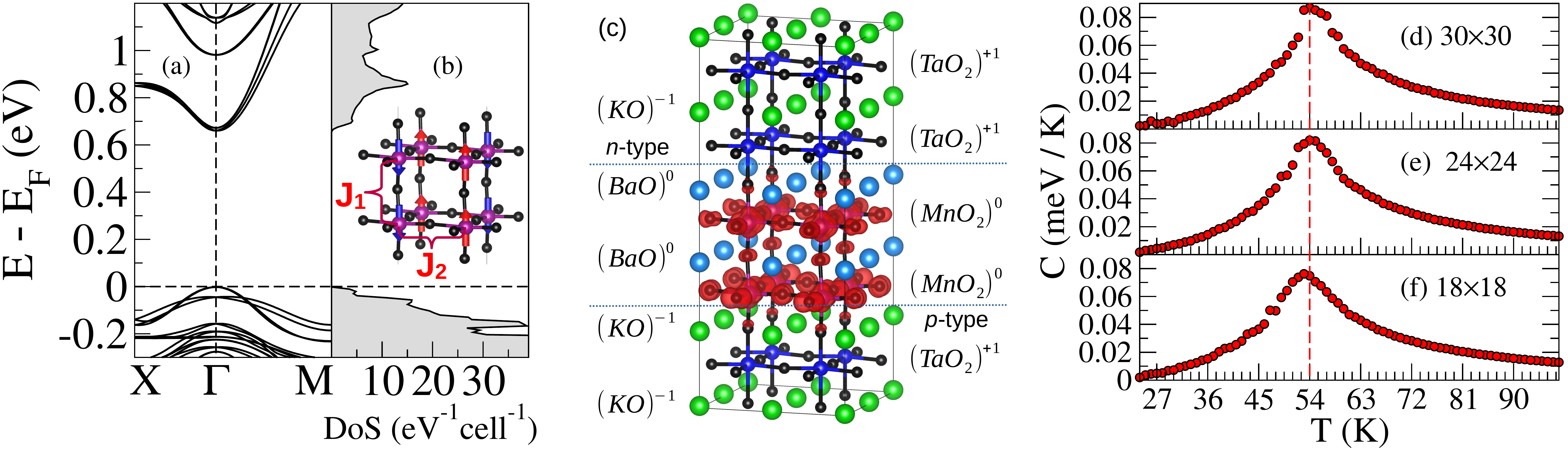}
\caption{\label{fig:DoSband}The band structure of (BaMnO$_3$)$_2|$(KTaO$_3$)$_3$ superlattice along $X \to \Gamma \to M$ direction including spin-orbit interaction for C-type antiferromagnetic arrangement with spin quantization along (001) direction is shown in (a), while (b) shows the corresponding density of states. The inset of (b) shows the BMO part of the heterostructure, marking the exchange paths for the interactions $J_1$ and $J_2$. (c) shows the partial charge density corresponding to the bands near the Fermi level, confirming their mapping onto the BaMnO$_3$ part. The {\scshape vesta} software has been used for preparing panel (c) \cite{VESTA}. The variation of magnetic specific heat as a function of temperature is shown for lattice sizes (d) $30 \times 30$, (e) $24 \times 24$, and (f) $18 \times 18$  within periodic boundary conditions.}
\end{figure*}
The band structure and the density of states (DoS) calculated within spin-orbit interaction, displayed in \cref{fig:DoSband}(a) and \cref{fig:DoSband}(b), indicate a gap of $\sim$0.65~eV between the valence and the conduction bands for the superlattice of our choice, with all bands close to the Fermi level (below and above) corresponding to the BMO part, as seen from the projected charge density in \cref{fig:DoSband}(c). The interface can be made a conducting one by increasing the KTO layer thickness, as indicated in \cref{eq:ChargeTransfer}. Our test calculation with (BaMnO$_3$)$_2|$(KTaO$_3$)$_{20}$ heterostructure reveals a conducting interface. Although the (BaMnO$_3$)$_2|$(KTaO$_3$)$_3$ heterostructure does not host a conducting interface desired for antiferromagnetic spintronics, we continue further calculations with this heterostructure, as we obtain invaluable insights about the spin-orbit interaction within an affordable computational cost that would be equally valid for a heterostructure with a conducting interface.

\begin{table}
	\caption{\label{tab:MagConfig}The energy difference per Mn atom, spin moment, and orbital moment at Mn sites obtained from our DFT calculations within LSDA + SOC + $U_\text{eff}$ = 3~eV for different magnetic arrangements and (001) spin-quantization axis are tabulated here.}
	\begin{ruledtabular}
	\begin{tabular}{c..}
		Arrangement & \multicolumn{1}{c}{Energy (meV)} & \multicolumn{1}{c}{Spin (orbital) moment ($\mu_{B}$)} \\
		\hline
        C-type AFM &  0.0 & 2.776 ~(0.028) \\
		FM         & 33.7 & 3.011 ~(0.027) \\
        A-type AFM & 34.5 & 2.981 ~(0.028) \\
		G-type AFM & 57.8 & 2.815 ~(0.028)
	\end{tabular}
	\end{ruledtabular}
\end{table}
\begin{table}
    \caption{\label{tab:MAE}The magneto-crystalline anisotropy energy per Mn atom calculated within LSDA + SOI + $U_\text{eff} = 3$~eV for bulk BMO (G-type antiferromagnet), BMO slab (A-type antiferromagnet), and BMO$|$KTO heterostructure (C-type antiferromagnet) are tabulated here for different spin quantization axes.}
    \begin{ruledtabular}
        \begin{tabular}{c...}
             Quantization & \multicolumn{3}{c}{Magneto-crystalline anisotropy energy (meV)} \\
             axis & \multicolumn{1}{c}{Bulk BMO} & \multicolumn{1}{c}{BMO slab} & \multicolumn{1}{c}{BMO$|$KTO} \\
             \hline
             100 & 0.000 & 0.000 & 0.123 \\
             010 & 0.000 & 0.001 & 0.125 \\
             110 & 0.000 & 0.001 & 0.000 \\
             001 & 0.000 & 0.074 & 0.080
        \end{tabular}
    \end{ruledtabular}
\end{table}
\begin{table}
    \caption{\label{tab:Exchange}The exchange interaction strengths estimated from our calculations are listed here.}
    \begin{ruledtabular}
        \begin{tabular}{c..}
            Exchange path & \multicolumn{1}{c}{Distance~(\AA)} & \multicolumn{1}{c}{Strength~(meV)} \\
            \hline
            $J_1$ & 3.77 & 4.96 \\
            $J_2$ & 3.99 & -1.86
        \end{tabular}
    \end{ruledtabular}
\end{table}
In order to find the lowest-energy magnetic configuration, we simulated A-type, C-type, and G-type antiferromagnetic (AFM) arrangements in the BaMnO$_3$ part, besides simulating the ferromagnetic (FM) arrangement. Our results, summarized in \cref{tab:MagConfig}, reveal the C-type antiferromagnetic arrangement to have the lowest energy. Further, our calculations considering spin-orbit interaction reveal that the system prefers a (110) spin quantization axis, making it the magneto-crystalline easy axis (see \cref{tab:MAE}). A comparison of the magneto-crystalline anisotropy energy (MAE) for the BMO$|$KTO heterostructure with that of bulk BMO in cubic form and a 2 unit cell thick BMO slab, as presented in \cref{tab:MAE}, indicates a small in-plane anisotropy for the BMO slab that gets reinforced upon heterostructuring with KTO due to proximity-induced spin-orbit interaction. We find magneto-crystalline isotropy for bulk cubic BMO.

The magnetic exchange interaction between the Mn atoms is mediated via the oxygen atoms, making superexchange the dominant mechanism of the magnetic interaction. We estimate the strengths of the magnetic exchange interactions by mapping our DFT results onto a spin Hamiltonian \cite{ChakrabortyPRB18}
\begin{equation}
	H = - \sum_{i,j} J_{ij} \vec{S}_i \cdot \vec{S}_j + \sum_i \epsilon_i^\text{an} \left| \vec{S}_i \right|^2, \label{eq:SpinHamiltonian}
\end{equation}
with $J$, $\vec{S}$, and $i$, $j$ representing the exchange interaction strength, spin vector and site indices, respectively. The second term describes the magnetic anisotropy, with $\epsilon_i^\text{an}$ being the anisotropy coefficient. Since the magneto-crystalline anisotropy energy is significantly weaker than the exchange interaction, we ignore the second term in \cref{eq:SpinHamiltonian} in the estimation of $J$. In the present case, examining various combinations of site indices, we find two dominant exchange interactions $-$ $J_1$ (out-of-plane) and $J_2$ (in-plane) (see \cref{fig:DoSband}(c)) $-$ whose values are determined by comparing the relative energies of the C-, G-, and A-type magnetic configurations and tabulated in \cref{tab:Exchange}.

The magnetic ordering temperature, also known as the N\'{e}el temperature in antiferromagnets, can be estimated using the exchange interaction strengths listed in \cref{tab:MagConfig}. The magnetic specific heat $C$ is estimated as \cite{RoojAPR23}
\begin{equation}
    C = \frac{1}{N} \frac{\partial U}{\partial T}=\frac{\left\langle \varepsilon^2 \right\rangle - \langle \varepsilon \rangle^2}{N k_B T^2},
\label{eq:specificHeat}
\end{equation}
where $\varepsilon$, $N$, $k_B$, and $T$ represent the energy of each magnetic configuration, the number of spins, the Boltzmann constant, and the temperature, respectively. $U(T) = \langle \varepsilon \rangle$ is the average internal energy. The energy of the magnetic configurations in the heterostructure with collinear spins and periodic boundary condition in $ab$-plane and only two unit cells along the $c$-direction is given by
\begin{equation}
    \varepsilon = -J_1 \sum_{i, j} S_i S_j - J_2 \sum_{i, j} S_i S_j,
\label{eq:Energy}
\end{equation} 
where $i, j$ iterate over the atoms satisfying the distance of the corresponding exchange path $J_1$ or $J_2$ (see \cref{fig:DoSband}(b) inset). In our calculations, we consider three different lattice sizes $30 \times 30, 24 \times 24$, and $18 \times 18$ within periodic boundary conditions to ensure that no finite lattice size effect corrupts our results. We employ the Metropolis Monte Carlo simulation algorithm to calculate $\left\langle \varepsilon^2 \right\rangle$ and $\langle \varepsilon \rangle$ at different temperatures. A spin $S_i$ is randomly chosen from the lattice, and the energy cost $\Delta \varepsilon$ for flipping the spin is computed. The flip is accepted if the corresponding Boltzmann weight, $\exp \left( -\Delta \varepsilon / k_B T \right)$, is greater than a random number $r \in(0,1)$. Another spin is randomly selected in the next time step. We average over $10^8$ imaginary time steps after bringing the system to thermal equilibrium in $10^9$ imaginary time steps for each temperature considered here. The results, shown in \cref{fig:DoSband}(d), \ref{fig:DoSband}(e), and \ref{fig:DoSband}(f) for the $18 \times 18, 24 \times 24$ and $30 \times 30$, respectively, within periodic boundary conditions, indicate an ordering temperature of $\sim$54~K $-$ a sizable value for an antiferromagnet, considering the fact that the structure is aperiodic along the $c$-direction. While the magnetic interactions are moderately short-ranged, the wide peak in the specific heat vs.\ temperature curve results from the mixture of ferromagnetic and antiferromagnetic interactions.

\begin{figure*}
    \centering
    \includegraphics[width = \textwidth]{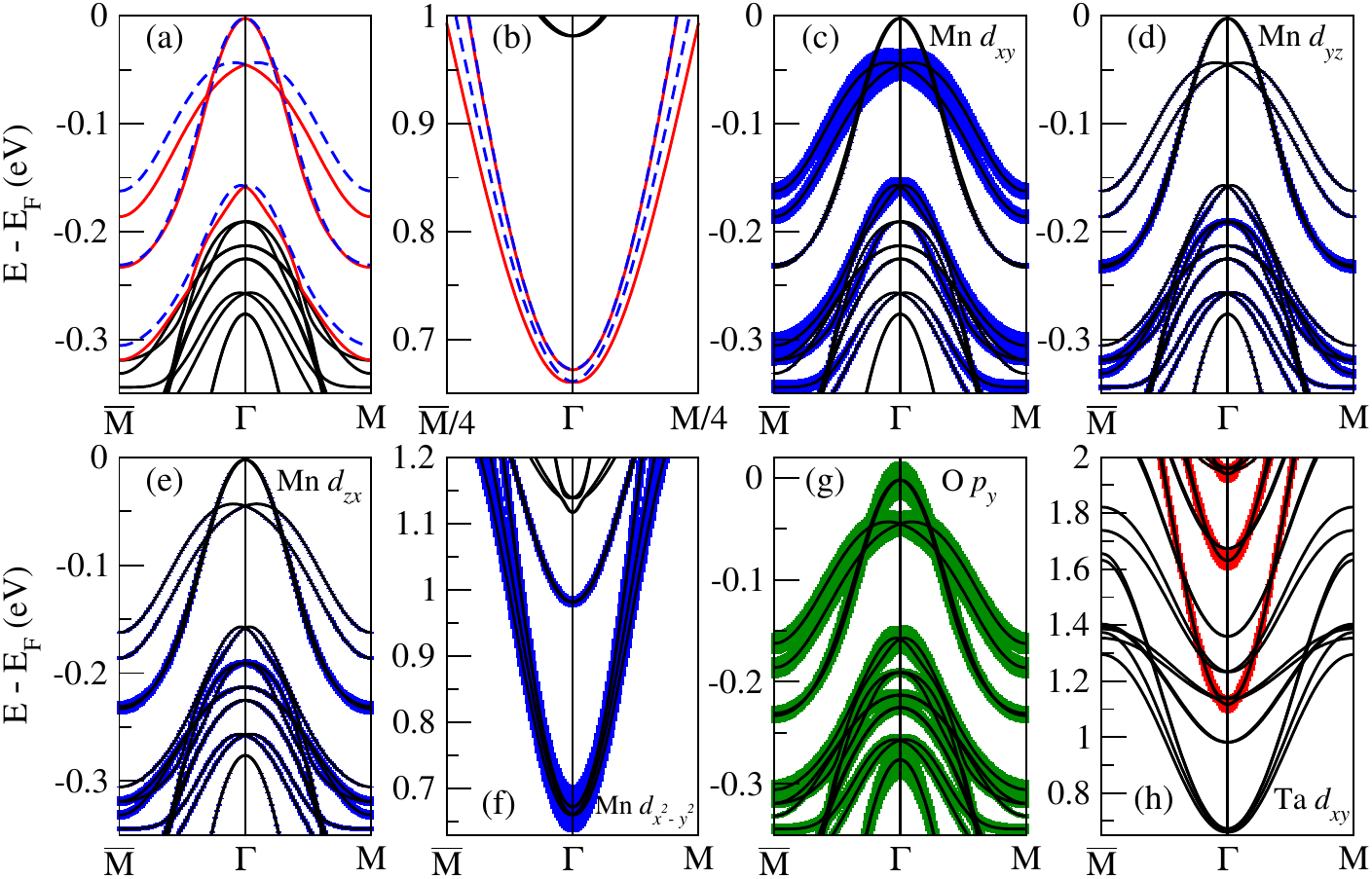}
    \caption{\label{fig:fatband}The spin-split bands of the (BaMnO$_3$)$_2|$(KTaO$_3$)$_3$ heterostructure along $\bar{M} \to \Gamma \to M$ below and above the Fermi level have been highlighted in (a) and (b), respectively, exhibiting Rashba splitting. A few pairs of the partner bands have been highlighted with solid red and dashed blue lines. The orbital characters of the Rashba-split bands near the Fermi level of the heterostructure are highlighted, with Mn $d_{xy}$, Mn $d_{yz}$, Mn $d_{zx}$, Mn $d_{x^2 - y^2}$, O $p_{y}$, and Ta $d_{xy}$ orbital characters featuring in (c), (d), (e), (f), (g), and (h), respectively.}
\end{figure*}

Although our calculation reveals no Rashba splitting for inversion-symmetric bulk BaMnO$_3$ in the cubic structure and a slab of BaMnO$_3$ with broken structure inversion symmetry, the band dispersion for the (BaMnO$_3$)$_2|$(KTaO$_3$)$_3$ heterostructure, displayed in \cref{fig:fatband}(a) and \cref{fig:fatband}(b), exhibits pronounced Rashba-like splitting of bands. A projected charge-density isosurface plot shown in \cref{fig:DoSband}(c) maps the Rashba-split bands near the Fermi level to the BaMnO$_3$ part of the heterostructure. While KTaO$_3$ is known to host strong spin-orbit interaction and a KTaO$_3$ slab exhibits strong Rashba splitting \cite{KumarPRB22}, our observation of pronounced Rashba-like splitting in the BaMnO$_3$ part suggests proximity-induced spin-orbit interaction at the BaMnO$_3$ part. We further analyze the orbital characters of the Rashba-split bands near the Fermi level,
as seen in \cref{fig:fatband}(c), \ref{fig:fatband}(d), \ref{fig:fatband}(e), and \ref{fig:fatband}(f) representing the orbital characters of Mn $d_{xy}$, Mn $d_{yz}$, Mn $d_{zx}$, Mn $d_{x^2 - y^2}$, O $p_{y}$, and Ta $d_{xy}$, respectively, suggesting a clear dominance of Mn $d$ orbitals in the valence and the conduction bands. The highly dispersive valence band maximum corresponds to O $p_y$ orbitals. The Ta-$d$ orbitals exhibit their presence at an elevated energy in the conduction bands, establishing a proximity-induced Rashba-like splitting in the BaMnO$_3$ part of the heterostructure.

After ascertaining that the proximity of Ta $5d$ orbitals with strong spin-orbit interaction induces substantial spin-orbit interaction in Mn $3d$ orbitals, we examine the exact nature of the Rashba-like interaction in the system.
\begin{figure*}
	\includegraphics[scale = 0.63]{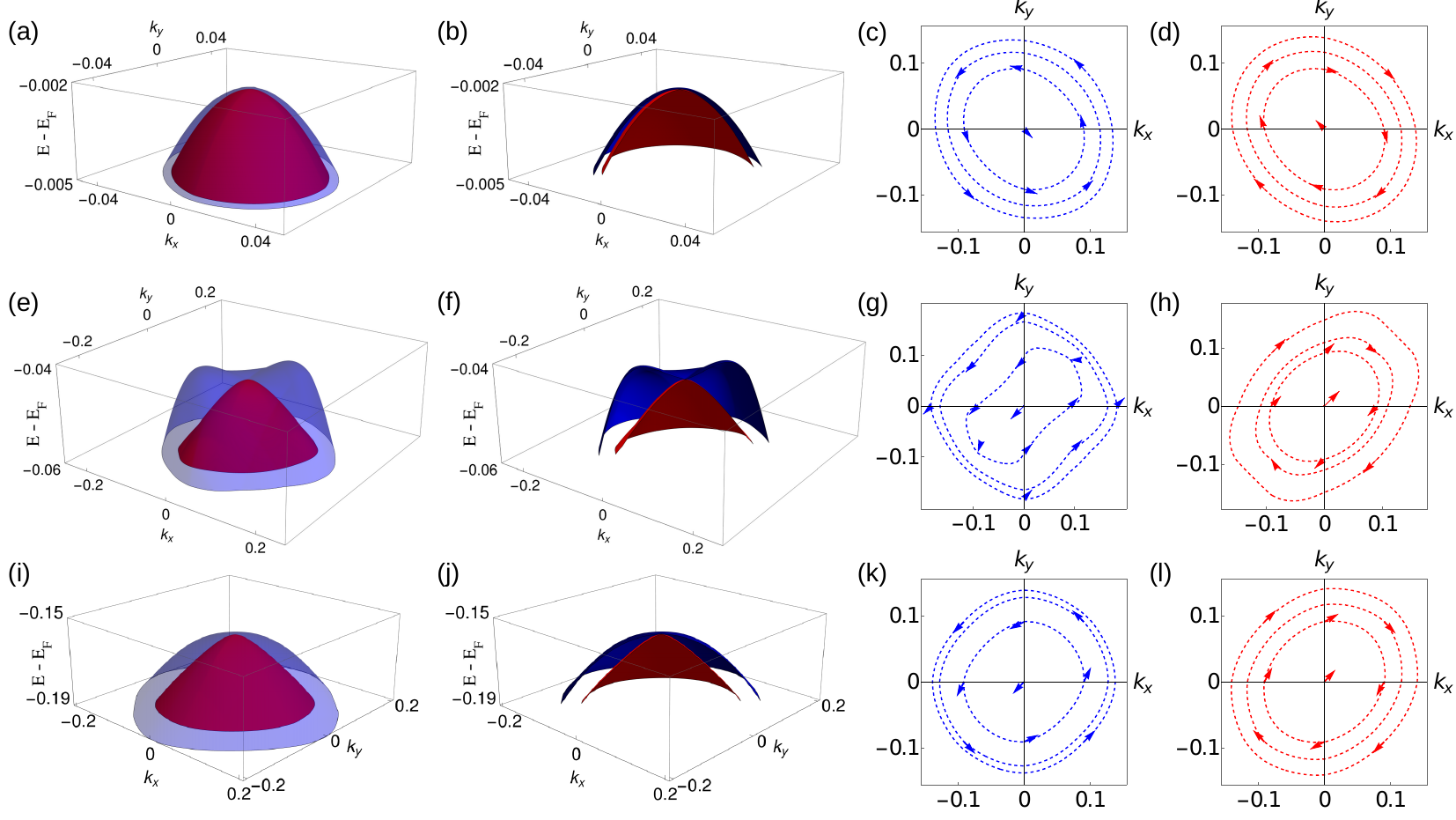}
	\caption{\label{fig:3dBands}The Rashba-like split bands near the Fermi energy are systematically visualized here with the help of three-dimensional (3D) bands, band cross sections, and spin textures. The 3D bands and their cross-section for the pair of bands at the valence band maximum (see \cref{fig:fatband}(a)) are visualized in (a) and (b), while the corresponding spin textures are shown in (c) and (d) on multiple isoenergetic contours. Similarly, the 3D bands and their cross-section for the pair of bands right below the valence band maximum at the $\Gamma$-point (see \cref{fig:fatband}(a)) are shown in (e) and (f), with the corresponding spin textures on multiple isoenergetic contours in (g) and (h). Finally, (i) and (j) show the 3D bands and their cross-section for the third highest pair of valence bands highlighted in \cref{fig:fatband}(a), with the corresponding spin textures on multiple isoenergetic contours in (k) and (l).}
\end{figure*}
Focusing on three pairs of bands near the valence band maximum, we plot the corresponding 3D band dispersion $E(k_x, k_y)$, their cross-sections, and the corresponding spin textures, where each spin vector is represented as $\vec{S} = \hat{x} \langle \psi | \hat{S}_x | \psi \rangle_{\vec{k}} + \hat{y} \langle \psi | \hat{S}_y | \psi \rangle_{\vec{k}}$, where $| \psi \rangle$ is the DFT-obtained electronic state, combined with isoenergetic contours, as displayed in \cref{fig:3dBands}. \Cref{fig:3dBands}(a) and \ref{fig:3dBands}(b) visualize the pair of bands and their cross-section, respectively, at the top of the valence bands near the $\Gamma$-point in 3D, clearly illustrating the splitting pattern. The corresponding helical spin textures, shown in \cref{fig:3dBands}(c) and \ref{fig:3dBands}(d), reveal the spins rotating along the bands' isoenegertic contours, resembling linear Rashba interaction \cite{KumarPRB22, GanguliPRB25}. The second and third pair of valence bands near the $\Gamma$-point exhibit a similar pattern of linear Rashba splitting and spin texture, as seen in \cref{fig:3dBands}(e-h) and \cref{fig:3dBands}(i-l), respectively. The second pair of bands obtained from our DFT calculations may be reasonably fitted to a two-band Rashba model having energy eigenvalues
\begin{equation}
    E_{R_{(1)}}^{\pm} = \frac{k^2}{2m^*} \pm \alpha_{(1)} k
\end{equation}
(see ref.~\cite{GanguliPRB25}), with the effective mass $m^* = -0.49m_e$ and the linear Rashba coefficient $\alpha_{(1)} = 0.10$~eV\AA, while the third pair fits to $m^* = -0.186m_e$ and $\alpha_{(1)} = 0.114$~eV\AA, $m_e$ being the electron's mass. Thus, our results show the induced spin-orbit interaction on the hybridized Mn and O bands near the Fermi level to be predominantly linear Rashba interaction. The negative signs in the effective mass indicate the usual convex upwards pattern of the valence bands near the $\Gamma$-point.

The above discussion calls for a comparison of the induced linear Rashba coefficient $\alpha_{(1)}$ from Mn bands with that of the KTaO$_3$ surface and LaAlO$_3|$SrTiO$_3$ heterostructure. The Rashba coefficient calculated by \citet{VarottoNC22} and \citet{ZhangACSN19} is approximately 0.31~eV\AA\ at KTaO$_3$ interfaces. The Rashba coefficient estimated here for the Mn-3$d_{xy}$ states in the BMO$|$KTO heterostructure is approximately four times larger than that found in the LaAlO$_{3}|$SrTiO$_{3}$ interface, as reported by \citet{LinNC19}, establishing the usefulness of proximity-induced SOI for the manifestation of moderate Rashba interaction in the BMO part of the BMO$|$KTO heterostructure. In this context, \citet{GuoCP25} reported an extremely high value of Rashba coefficient in the valence and conduction bands of antiferromagnetic MnBi$_2$Te$_4$, in excess of 4~eV\AA, arising from Bi-$p$ and Te-$p$ bands, not from the magnetic Mn bands. Since our proposal for antiferromagnetic spintronics relies on the tunability of magnetic domains via spin-orbit torque, MnBi$_2$Te$_4$ is not relevant in the present context. While larger values of the Rashba coefficient has been found in many other systems, the sizable proximity-induced Rashba interaction observed here in the antiferromagnetic Mn-$3d$ bands, is unprecedented. Thus, our results present an interesting opportunity for antiferromagnetic spintronics via proximity-induced spin-orbit interaction. We also note that since the Ta bands are nowhere close to the Fermi level, they do not interfere with the Rashba-split bands near the Fermi level.
\section{\label{sec:conc}Conclusion}
In summary, we simulate a heterostructure of BaMnO$_3|$KTaO$_3$ with the view of combining a robust antiferromagnet and a strong spin-orbit interaction material to realize a heterostructure useful for antiferromagnetic spintronics. The electrostatic potential developed in the superlattice due to a combination of polar$|$nonpolar materials result in BMO bands near the Fermi level and an onset of charge transfer beyond 10 unit cell thick KTO layer. The BMO part of the heterostructure shows strong antiferromagnetism, with a significant magnetic moment at the Mn sites and reasonably strong interactions, leading to an ordering temperature of at least 54~K. Our results confirm the induction of a moderate spin-orbit interaction in the Mn bands of the BMO part due to its proximity to the KTO part, leading to a substantial linear Rashba coefficient. Thus, based on our estimation of electrostatic potential, DFT calculations, and Monte Carlo simulations, we demonstrate the BMO$|$KTO heterostructure to offer a combination of substantial antiferromagnetism and a moderate Rashba spin-orbit interaction $-$ a combination of promising features for antiferromagnetic spintronics $-$ owing to the proximity effect. Besides, the idea of designing heterostructures for proximity-induced features can motivate future theoretical and experimental works.

\begin{acknowledgments}
Fruitful scientific discussion with Dr.\ Jayita Chakraborty, financial support from SERB, India, through grant numbers CRG/2021/005320 and ECR/2016/001004, and the use of a high-performance computing facility at IISER Bhopal are gratefully acknowledged.
\end{acknowledgments}

%
\end{document}